\documentclass[preprint, 5pt]{elsarticle}

\usepackage{boites}
\usepackage{amsmath,amssymb}
\usepackage{lscape}
\usepackage{subfigure}

\begin{document}

\newcommand{\etal}      {{\it et~al.}}

\title{First Millimeter-wave Spectroscopy of the Ground-state Positronium}

\author{A. Miyazaki,$^1$ T. Yamazaki,$^1$ T. Suehara,$^2$ T. Namba,$^1$ S. Asai,$^1$ T. Kobayashi,$^1$ H. Saito,$^3$ Y. Tatematsu,$^4$ I. Ogawa,$^4$ and T. Idehara,$^4$}

\address{$^1$Department of Physics, Graduate School of Science, and International Center for Elementary Particle Physics, University of Tokyo, 7-3-1 Hongo, Bunkyo-ku, Tokyo 113-0033, Japan \\$^2$Department of Physics, Faculty of Science, Kyushu University, 6-10-1 Hakozaki, Higashi-ku, Fukuoka 812-8581, Japan\\$^3$Institute of Physics, Graduate School of Arts and Sciences, University of Tokyo, 3-8-1 Komaba, Meguro-ku, Tokyo 153-8902, Japan \\$^4$Research Center for Development of Far-Infrared Region, University of Fukui, 3-9-1 Bunkyo, Fukui-shi, Fukui 910-8507, Japan}

\begin{abstract}%
We report on the first measurement of the Breit-Wigner resonance of the transition from {\it ortho-}positronium to {\it para-}positronium.
We have developed an optical system to accumulate a power of over 20~kW using a frequency-tunable gyrotron and a Fabry-P\'{e}rot cavity.
This system opens a new era of millimeter-wave spectroscopy, 
and enables us to directly determine both the hyperfine interval and the decay width of {\it p-}Ps.
\end{abstract}


\parindent0pt

\maketitle

\section{Introduction}
Positronium (Ps)~\cite{Ps-REV} is a bound state of an electron and a positron.
Ground-state positronium has two spin eigenstates:
{\it ortho-}positronium~({\it o-}Ps, Spin $=1$, $3\gamma$-decay, lifetime = 142~ns~\cite{LIFE-ORTH}) 
and {\it para-}positronium~({\it p-}Ps, Spin $=0$, $2\gamma$-decay, lifetime = 125~ps~\cite{LIFE-PARA}).
The energy level of {\it o}-Ps is higher than that of {\it p}-Ps by the hyperfine structure ($\Delta^{\rm Ps}_{\rm HFS}$).
Compared with the hyperfine structure of hydrogen (about 1.4~GHz),
$\Delta^{\rm Ps}_{\rm HFS}$ is very large, about 203~GHz (wavelength $=1.5$~mm),
due to light mass of Ps and an s-channel contribution (87~GHz). 
Since the transition from {\it o-}Ps to {\it p-}Ps is forbidden, 
high-power (over 10~kW) millimeter-wave radiation is required to measure the resonance around the hyperfine structure.
Many technological difficulties regarding the use of millimeter waves have prevented a direct measurement of this resonance.
Measurements using the Zeeman effect in a static magnetic field ($\sim$1~T) have been intensively studied instead of direct measurements.
However, it has been strongly desired to directly examine the Breit-Wigner resonance from free {\it o-}Ps to {\it p-}Ps
because properties of Ps is derived in a fundamental way by Quantum Electrodynamics without any external fields.

In this paper, we present the first results of the measurement of the resonance transition in ground-state free Ps.
We developed a very challenging system of high-power and frequency tunable millimeter-wave devices for this measurement.
As a result of the measurement of the Breit-Wigner resonance,
we can directly determine both $\Delta^{\rm Ps}_{\rm HFS}$ and the decay width of {\it p-}Ps ($\Gamma_{\rm p\text{--}Ps}$).
It should be noted that determination of these two values is the first achievement for free Ps.
The present work is the first demonstration of spectroscopy by scanning the frequency of high-power millimeter waves.
This new method also paves the way for various measurements in material and life science, 
such as the DNP-NMR spectroscopy~\cite{DNP-NMR}.

\section{Experimental setup}
\subsection{Technical achievements}
\begin{figure}[h]
\begin{center}
\includegraphics[width=90mm, angle=-90]{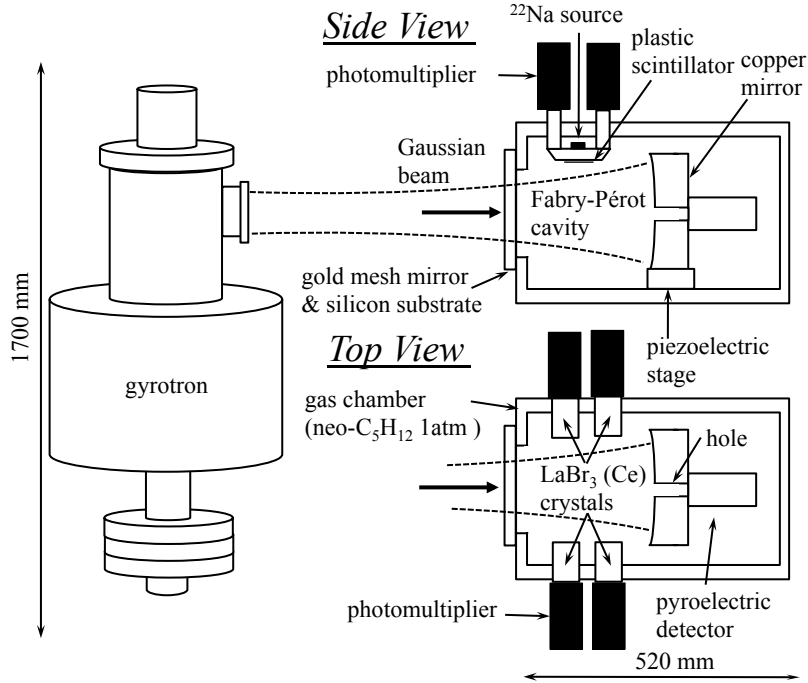}
\caption{
Schematic view of the experimental setup.
Top and side views of the gas chamber are shown.
\label{fig:setup1}
}
\end{center}
\end{figure}

The spectroscopy of the transition from {\it o-}Ps to {\it p-}Ps requires frequency-tunable (201--205~GHz) and high-power (over 20~kW) millimeter waves.
Previously~\cite{HFS-DIRECT}, a millimeter-wave source, gyrotron, was totally monochromatic (202.89~GHz) and its output profile is an impure Gaussian (about 30\%).
A Fabry-P\'erot cavity which accumulates millimeter waves from the gyrotron was unable to store over 11~kW.
We developed two innovative devices in the millimeter-wave range:
\begin{enumerate}
\item A frequency tunable gyrotron with output of a Gaussian profile (purity is over 95\%).
\item A high-gain Fabry-P\'erot cavity withstanding very high power.
\end{enumerate}
Figure \ref{fig:setup1} shows the apparatus of our setup.
The Fabry-P\'erot cavity is placed inside a gas chamber which will be described later.

The gyrotron is a cyclotron-resonance-maser fast wave device, whose output power ($P_g$) is highest ($>100$~W) in the millimeter-wave range.
An electron beam gyrating in a strong magnetic field ($\sim 7$~T) bunches to a deceleration phase 
and excites a resonant mode (millimeter waves) of a cavity in the gyrotron.
We have successfully developed a new gyrotron (FU CW GI) operating in the TE$_{52}$ mode with an internal mode converter~\cite{FU-CW-G1}.
This gyrotron works in pulsed operation (duty ratio 30\%, repetition rate 5~Hz),
with which all data are acquired in synchronization
 (events collected during and outside gyrotron pulses are defined as beam-ON and beam-OFF, respectively).
The output millimeter-wave beam has a Gaussian profile.
The electron beam current is monitored and fed back to control the voltage of the heater of the gyrotron's electron gun.
The power of the output beam was thus stabilized to within $\sim$10~\% during each measurement (lasting a few days).

The frequency of the gyrotron is tuned between 201~GHz and 205~GHz by using gyrotron cavities of different radius ($R_0$).
The values of the frequency and cavity radius are summarized in Table~\ref{tab:G1_used}.
A far off-resonance point ($180.56$~GHz) is obtained by using a different operating mode (TE$_{42}$ mode).
When the cavity of 2.467~mm was used,
we changed the strength of gyrotron's magnetic field so that oscillation frequency moved within Q-value of the cavity.
The frequency is precisely measured ($\pm$1~kHz) using a heterodyne detector (Virginia Diodes Inc., WR5.1 Even Harmonic Mixer).
Using this method, we have successfully overcome many difficulties in tuning high-power millimeter waves, which are basically monochromatic.

\begin{table}[tbp]
\begin{center}
  \caption{Operating points.\label{tab:G1_used}}
    \begin{tabular}{ccccc}
    \hline
    $R_0$ [mm]  & mode      & frequency  [GHz] & $P_g$ [W] & $P_{\rm eff}$ [kW]\\ \hline
    2.453       & TE$_{42}$ & 180.56           & 300       & 41 \\
    2.481       & TE$_{52}$ & 201.83           & 190       & 22 \\
    2.475       & TE$_{52}$ & 202.64           & 240       & 23 \\
    2.467       & TE$_{52}$ & 203.00           & 550       & 21 \\
    2.467       & TE$_{52}$ & 203.25           & 250       & 25 \\
    2.463       & TE$_{52}$ & 203.51           & 350       & 30 \\
    2.453       & TE$_{52}$ & 204.56           & 410       & 25 \\
    2.443       & TE$_{52}$ & 205.31           & 125       & 24 \\ \hline
    \end{tabular}
\end{center}
\end{table}%

As shown in Fig.~\ref{fig:setup1}, the beam from the gyrotron is guided into the Fabry-P\'erot cavity, 
which consists of a gold mesh plane mirror (diameter~$= 50$~mm, line width~$= 200$~$\mu$m, separation~$= 140$~$\mu$m, thickness~$= 1$~$\mu$m) and a copper concave mirror (diameter~$= 80$~mm, curvature $= 300$~mm, reflectivity $= 99.85$\%).
The cavity length (156~mm) is precisely controlled ($\sim$100~nm) by a piezoelectric stage under the copper mirror (side view of Fig.~\ref{fig:setup1}).
The accumulated power in the cavity is measured using the radiation transmitted through a hole (diameter~$= 0.6$~mm) at the center of the copper 
mirror.
This transmitted radiation is monitored by a pyroelectric detector.

The gold mesh mirror is fabricated on a high-resistivity silicon plate (thickness~$= 1.96$~mm).
This silicon substrate, blocking optical photons,  is also used as the window of the gas chamber.
The use of silicon as a base is a technical breakthrough to withstand at most 80~kW effective power with water cooling.
Thanks to this high effective power, we can obtain enough signals of the transition from {\it o-}Ps to {\it p-}Ps to determine the resonance shape.
One disadvantage of the silicon is a severe interference of millimeter waves between the mesh mirror and the silicon plate due to its high refractive index~($3.45$).
In order to reduce this effect, CST Microwave Studio~\cite{CST} is used to simulate the interference and to optimize the structure of the mesh mirror.
A high reflectivity ($\sim $99.1~\%) and low loss ($\sim$ 0.3~\%) are obtained at frequencies around 203~GHz.
The power stored in the Fabry-P\'{e}rot cavity ($P_{\rm eff}$) is designed to be over 20~kW when the power of input radiation is over 100~W.

\subsection{Power estimation}
\begin{figure}[h]
\begin{center}
\includegraphics[width=90mm, angle=-90]{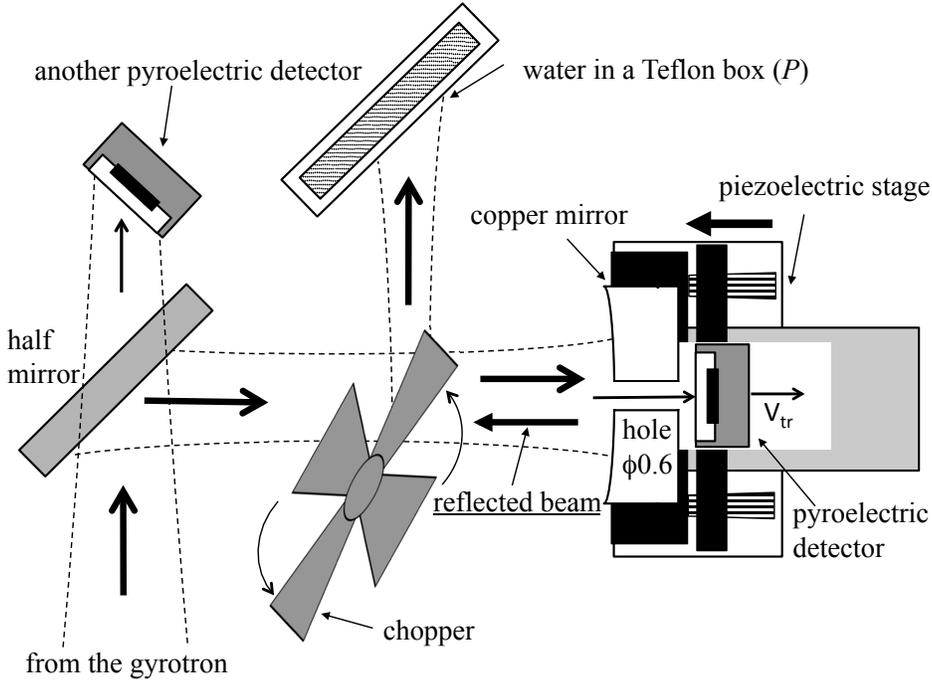}
\caption{
Schematic view of the setup used to estimate the absolute accumulated power.
\label{fig:setup2}
}
\end{center}
\end{figure}

Absolute power estimation of high-power millimeter waves is very difficult.
Moreover, we should calibrate the power stored inside the Fabry-P\'erot cavity.
The absolute accumulated power is measured as shown in Fig.~\ref{fig:setup2}.
The ratio between the accumulated power and the radiation transmitted through the hole on the copper mirror is calibrated using the beam from the gyrotron.
A chopper splits the beam in order to simultaneously measure the transmitted signal and the beam power.
This reduces systematic uncertainties due to time-dependent (a few minutes) instability of the gyrotron output.
The chopper is synchronized to the gyrotron pulses, and switches the propagation direction from one pulse to the next.
Half of the pulses are totally absorbed in a Teflon box filled with water (46~ml).
Transparency of Teflon was measured in advance (95\% $\pm$ 5\%).
Heat dissipation from water was corrected by fitting cooling curves by theory, and obtained cooling rates were confirmed by additional measurements.
The power $P$ is estimated by a temperature increase  of the water.
The other half are passed to the copper mirror, where the power transmitted through its hole is measured by the pyroelectric detector (output voltage~$=V_{\rm tr}$).
The calibration factor $C$ is defined as $C\equiv 2P/V_{\rm tr}$~[kW/V].
The factor 2 comes from back-and-forth waves in the Fabry-P\'erot cavity.

Using this method, the accumulated power $P_{\rm eff}=CV_{\rm tr}$ is measured (Table~\ref{tab:G1_used}).
The stored power is always over 20~kW, which is twice as high as previous power (11~kW).
The result is consistent with a rough estimated $P_{\rm eff}$ considering the finesse~(400-600) and coupling~(about 60\%) of the cavity~\cite{OPTICS}.
To control  $P_{\rm eff}$ for all frequency points,
we placed a wire grid polarizer between the gyrotron and the Fabry-P\'erot cavity.
We measure $C$ before and after the transition measurement at each frequency; since these two results are consistent, their mean value is used in the analysis.

\subsection{Formation of positronium}
Positronium is formed in the gas chamber in which the Fabry-P\'erot cavity is placed (Fig.~\ref{fig:setup1}).
A positron emitted from a $^{22}$Na source (1~MBq) is tagged by a thin plastic scintillator (thickness~$=0.1$~mm, NE-102 equivalent), 
and the $\gamma$ rays produced in its annihilation are detected by four LaBr$_{3}$(Ce) crystals.
Time spectra of Ps is obtained as a time difference between the positron and $\gamma$-ray signals.
Photomultipliers (HAMAMATSU R5924-70) are used to detect optical photons from the scintillators, and charge-sensitive ADCs are used to measure the energy.
The temperature of the chamber is maintained at less than $30 ^{\circ}$C using water-cooling.

\begin{figure}[h]
\centering
\subfigure[]{\includegraphics[width=120mm]{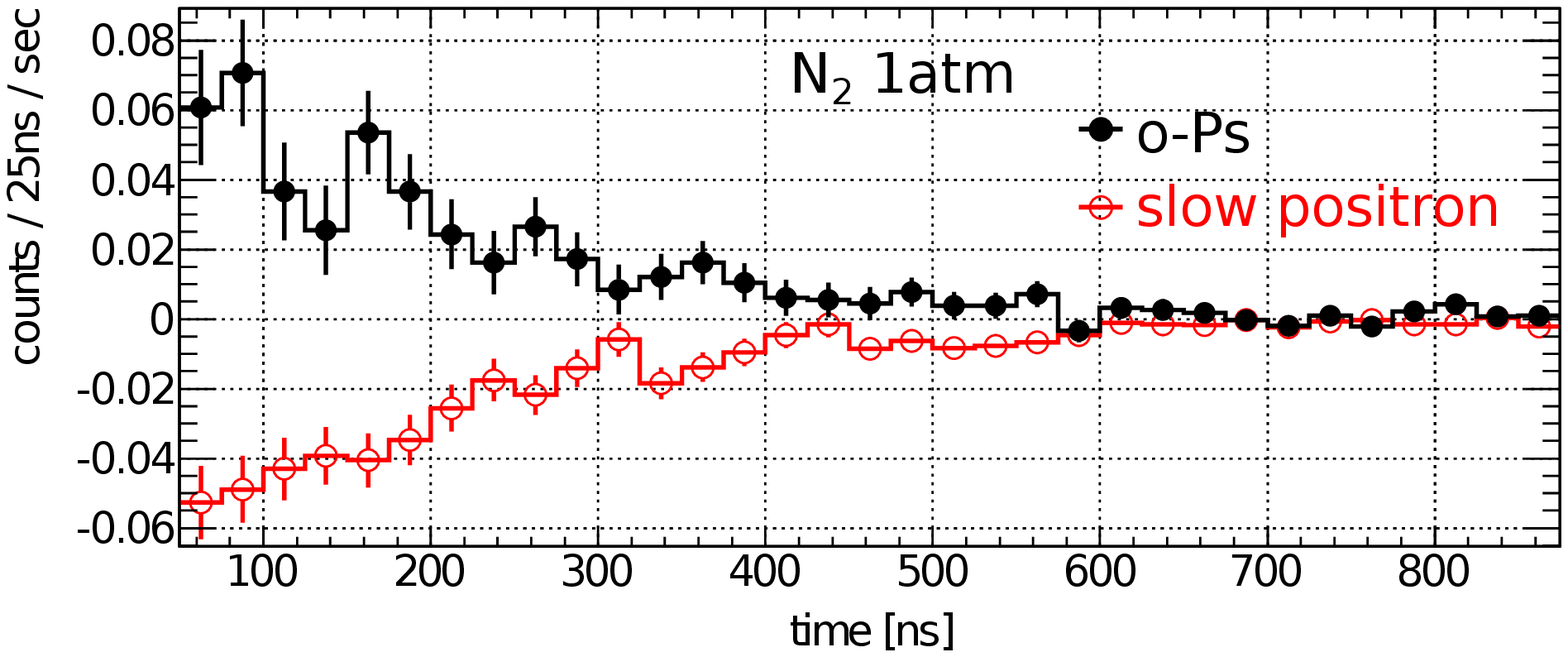}}
\\
\centering
\subfigure[]{\includegraphics[width=120mm]{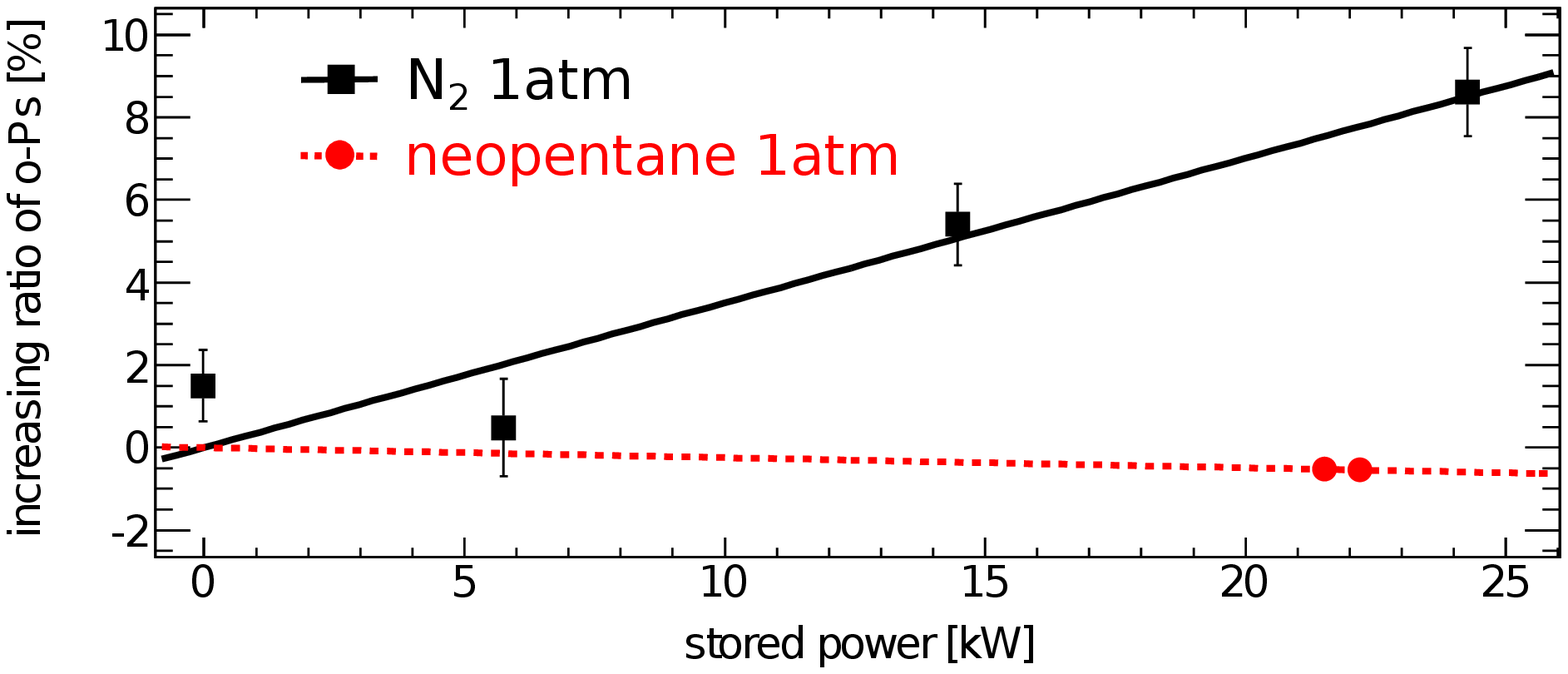}}
\caption{
(a) Time spectra subtracted beam-OFF events from beam-ON ones in which frequency is 201.8~GHz and stored power is 24~kW.
Slow positron (selected by back-to-back $\gamma$ rays of 511~keV) and {\it o-}Ps ($\gamma$ rays from 340~keV to 450~keV) in pure N$_2$ (1~atm) are shown.
(b) Power dependence of increasing ratio of {\it o}-Ps in N$_2$ gas and neopentane.
A solid line shows a linear fit for {\it o-}Ps data and a dashed line does that for neopentane.
The origin of neopentane data is defined as zero due to lack of data at low power.
\label{fig:psform}}
\end{figure}

There has been a long-standing problem of increasing ratio of Ps production in gas when irradiated by electromagnetic waves.
It was firstly reported by Ref.~\cite{HFS-FIRST},
and was studied in a static electric field using the Boltzmann equation~\cite{PS-FORM}.
We have also observed the same phenomenon using millimeter waves~\cite{HFS-DIRECT}.
In order to further investigate this phenomenon, we measured time spectra of {\it o-}Ps and slow positrons (positron with energy below threshold for Ps production~\cite{SLOW-POS}) in pure N$_2$.
The spectra in N$_2$ shown in Fig.~\ref{fig:psform}(a) clearly demonstrates the increase of {\it o}-Ps and decrease of slow positrons.
This phenomenon is due to a slow positron accelerated by the strong millimeter-wave fields~\cite{PHD-MIYAZAKI}.
The accelerated slow positrons collide randomly with gas molecules with a rate comparable to 203~GHz and finally exceed the threshold of Ps production (Ore gap).
Figure~\ref{fig:psform}(b) shows that increasing ratio of Ps is almost proportional to the stored power.
This phenomenon does not strongly depend on frequency and causes fake signals at off-resonance points;
therefore it would distort the Breit-Wigner resonance by a few \% level.

The cause of this phenomenon is an elastic scattering of slow positrons with N$_2$ gas molecules.
Target gas is required to have many vibrational and rotational modes
because its cross-section of inelastic scattering is large and drastically decelerates the accelerated slow positrons as indicated in Ref.~\cite{PS-FORM}.
We selected pure neopentane (C-(CH$_{3}$)$_{4}$) gas (25$^\circ$, 1~atm) which has much more internal degrees of freedom than N$_2$.
No increase of {\it o}-Ps in neopentane (Fig.~\ref{fig:psform}(b)) justifies our hypothesis.
The use of neopentane also provides high stopping power and efficient Ps production.
Furthermore, neopentane does not absorb millimeter waves, 
differently from isobutane (mixed to N$_2$ in Ref.~\cite{HFS-DIRECT}) which has an absorption line at 202.5~GHz~\cite{ROTATION}.
To confirm that the use of neopentane surely eliminates the problem in Ps production,
a far off-resonance point ($180.56$~GHz) was measured.


\section{Analysis}
To enhance {\it o}-Ps events,
we require that the time difference between the plastic scintillator signal and the coincidence signal of the LaBr$_3$(Ce) detectors is between 50~ns and 250~ns.
Pileup events in which two different positrons signal the plastic scintillator are reduced 
by requiring that the charge measured by long (1000~ns) and short (60~ns) gate ADCs are consistent.
The number of accidental coincidence is estimated using the time window between 850~ns and 900~ns, and is subtracted from the signal sample.
We also apply an energy selection, between 494~keV and 536~keV, to select the two back-to-back $\gamma$ rays.

\begin{figure}[h]
\centering
\includegraphics[width=100mm]{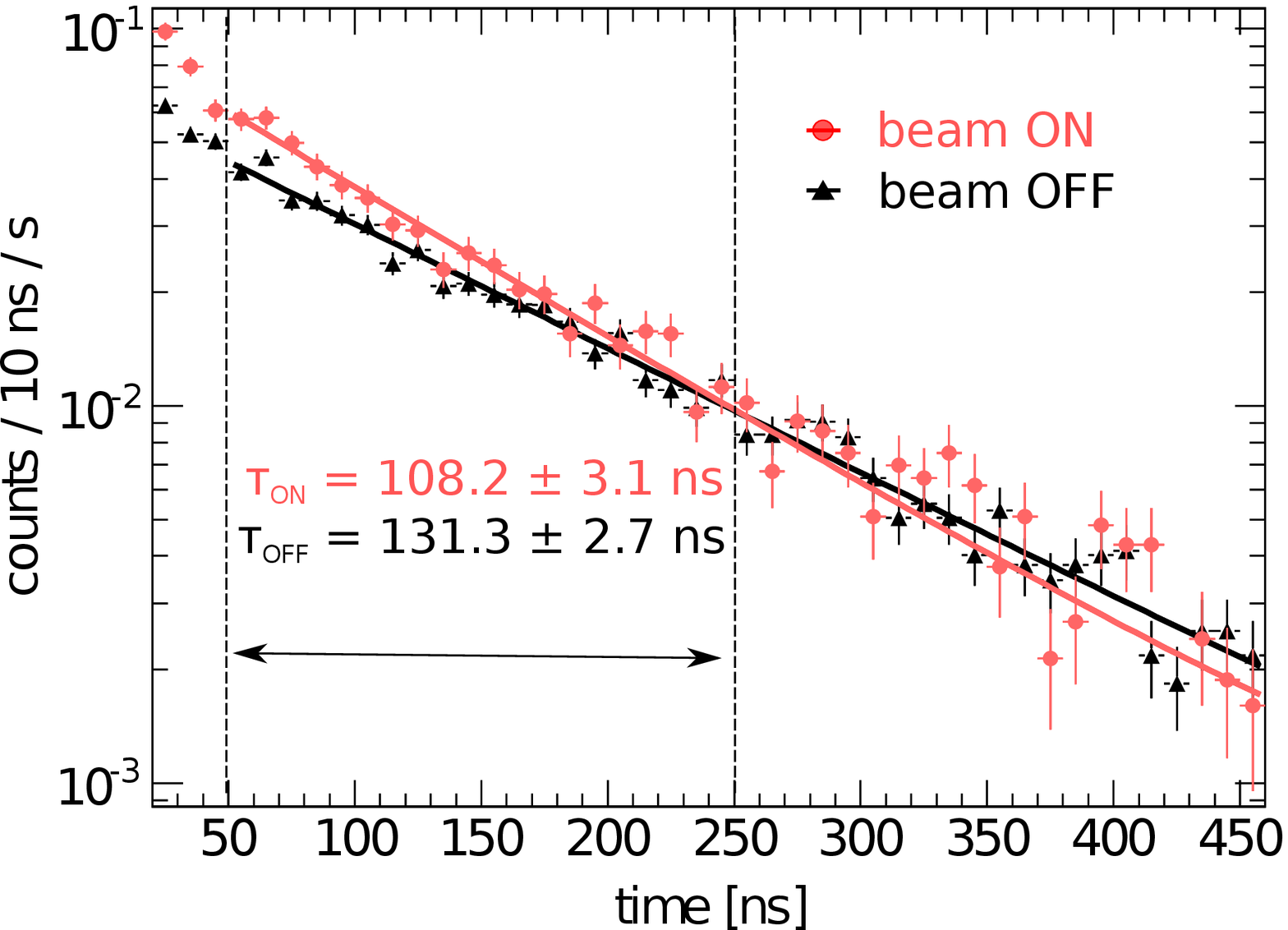}
\caption{
(color on-line)
Time spectra of the LaBr$_3$(Ce) scintillator at $203.51$~GHz and $67.4$~kW,
after the rejection of accidental events and the energy selection.
The solid lines show the results of fits to exponential functions.
The chosen time window is shown by the two dashed lines.
\label{fig:tspec}}
\end{figure}

Figure~\ref{fig:tspec} shows the measured time spectra at a frequency of $203.51$~GHz and accumulated power of $67.4$~kW.
Data are shown separately for events in beam-ON or beam-OFF of gyrotron pulses.
The beam-OFF spectrum consists of pick-off annihilation (quenching by an electron in a gas molecule~\cite{PICK-OFF}) and $3\gamma$-decays of {\it o}-Ps.
The lifetime shortened by the transition from {\it o-}Ps to {\it p-}Ps
($\tau_{\rm OFF}=131.3\pm2.7\, {\rm ns} \rightarrow \tau_{\rm ON}=108.2\pm3.1\, {\rm ns}$) is observed as shown in Fig.~\ref{fig:tspec}.
This decrease in lifetime is consistent with the theoretical prediction, 
and results in an enhancement of the event rate during the beam-ON period.
The event rates in beam-ON and beam-OFF periods are $R_{\text{ON}}=548$~mHz and $R_{\text{OFF}}=455$~mHz, respectively.

The reaction cross-section $\sigma$ of the transition from {\it o-}Ps to {\it p-}Ps is obtained
by comparing the measured $S/N \equiv (R_{\text{ON}}-R_{\text{OFF}})/R_{\text{OFF}}$ with the value simulated using the stored power.
The calibrated effective power in the Fabry-P\'erot cavity is continuously monitored
by measuring the $V_{\rm tr}$ waveform using a sampling ADC (sampling rate of 0.5~kHz).
We estimate the position of Ps formation and relative detection efficiencies of 2$\gamma$- and 3$\gamma$-decays using GEANT4 simulation~\cite{GEANT4}.
The transition probability is calculated using the Ps positions and the theoretical distribution of the electromagnetic field within the cavity.
We then obtain the relation between $S/N$ and $\sigma$, and numerically solve the equation $S/N (\sigma) = (R_{\text{ON}}-R_{\text{OFF}})/R_{\text{OFF}}$.
The advantage of using $S/N$ is that the least well constrained parameters used in the simulation 
(absolute source intensity, detector misalignment, and stopping position of positrons) are canceled out.
We also measure $S/N$ when the Fabry-P\'erot cavity does not accumulate millimeter waves, in which case $S/N$ is consistent with zero.

\section{Result}
\begin{figure}[h]
\centering
\includegraphics[width=100mm]{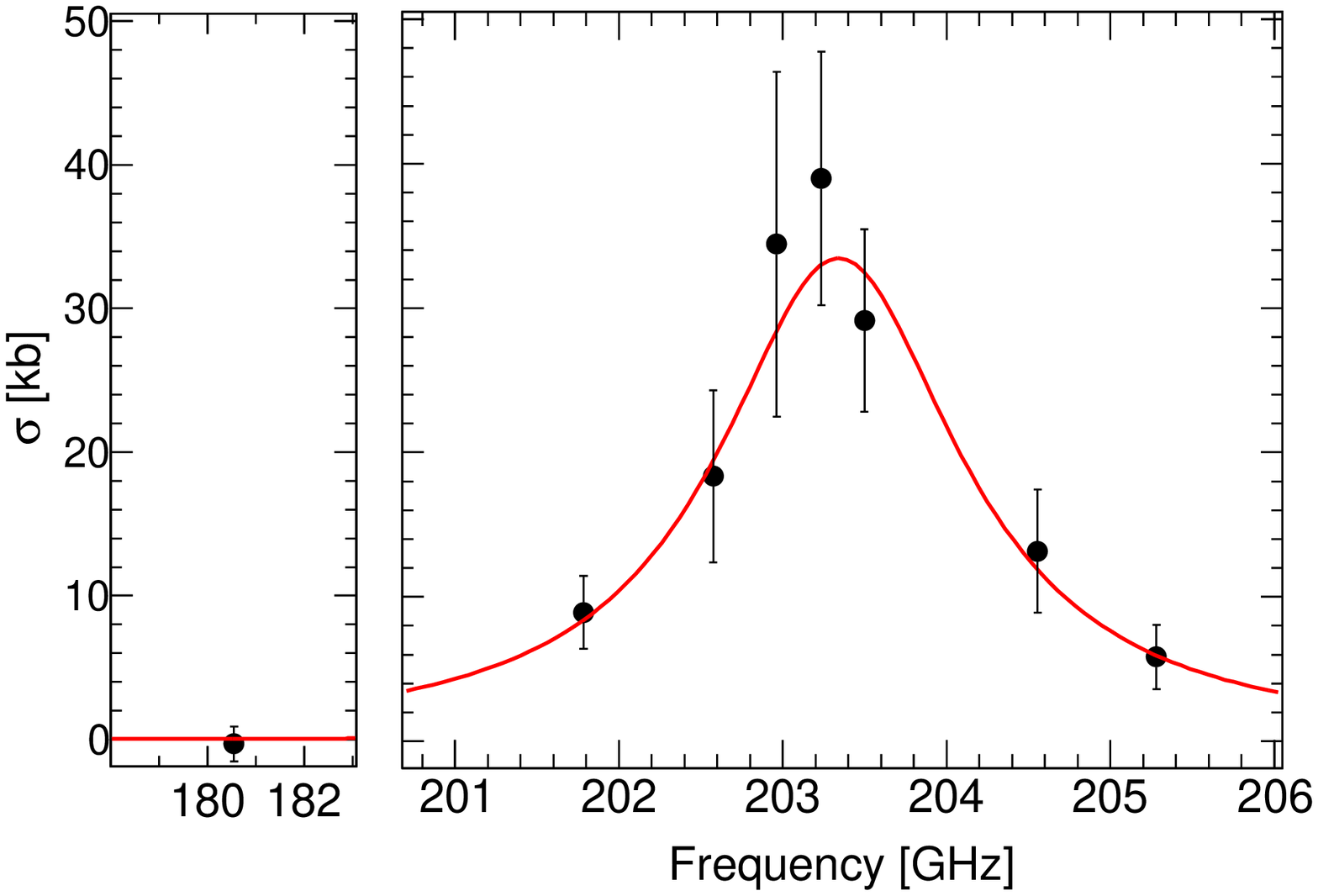}
\caption{
Measured reaction cross-section of the direct transition.
The solid line is the best fit (using only statistical errors) to a Breit-Wigner function.
\label{fig:reso}}
\end{figure}

Figure~\ref{fig:reso} shows the obtained result of cross-sections versus frequency.
Data at far off-resonance ($180.56$~GHz) demonstrate the absence of fake signals.
A clear resonance is obtained.
The data are fitted by a Breit-Wigner function of the angular frequency $\omega$
\begin{equation}
g(\omega) = 3A \frac{\pi c^2}{\hbar^2\omega_0^2} \cdot \frac{1}{\pi} \frac{\Gamma/2}{(\omega-\omega_0)^2 + (\Gamma/2)^2},
\end{equation}
where $\omega_0$ is $2\pi\Delta^{\rm Ps}_{\rm HFS}$, $A$ is the Einstein A coefficient of this transition, 
and $\Gamma$ is the natural width of the transition.
Using the decay width of {\it o-}Ps ($\Gamma_{\rm o\text{--}Ps}$) and $\Gamma_{\rm p\text{--}Ps}$,
$\Gamma$ is expressed by
\begin{equation}
\Gamma = A + \Gamma_{\rm p\text{--}Ps} + \Gamma_{\rm o\text{--}Ps}.
\end{equation}
Since $A$ and $\Gamma_{\rm o\text{--}Ps}$ are much smaller than $\Gamma_{\rm p\text{--}Ps}$, 
$\Gamma$ is approximated by $\Gamma_{\rm p\text{--}Ps}$.
We therefore treat $\Delta^{\rm Ps}_{\rm HFS}$, $\Gamma_{\rm p\text{--}Ps}$ and $A$ 
as the three parameters to be determined in the fit.

Systematic errors are summarized in Table~\ref{tab:sys}.
The second largest systematic is about the power calibration factor $C$.
The systematic error on $C$ is from the measurement of the water temperature ($10$\%) and correction of the spatial distribution ($10$\%).
This was combined with the variations of $C$ observed under different reflection conditions.
The standard deviation of this fluctuation is between $9$\% and $20$\% for the different gyrotron cavities.
At each frequency, we propagate uncertainty of $C$ to the three fitting parameters.

The Stark effect due to the electric field of gas molecules induces a shift in $\Delta^{\rm Ps}_{\rm HFS}$.
This effect is estimated from the measurements in N$_2$ gas used in Ref.~\cite{HFS-EXP},
assuming that it depends linearly on the number of density and the scattering cross-section obtained in Doppler-broadening measurements~\cite{DOPPLER}.
The shift is corrected ($+460$~ppm) and the amount of this correction is conservatively assigned as a systematic error.
A linear extrapolation is sufficient at the current experimental precision,
however, as has recently been pointed out~\cite{THERM-OKINAWA}\cite{HFS-PREC},
the effect of non-thermalized Ps distorts the linearity by around 10--20 ppm, 
and may be problematic for more precise measurements.

We also estimate an uncertainty due to detection efficiencies obtained using GEANT4 simulation.
Since $S/N$ is used to obtain the cross-sections, 
only the relative efficiency between 2$\gamma$- and 3$\gamma$-decays takes part in the uncertainty.
Energy spectra of beam-OFF events are fitted with the simulated spectra of 2$\gamma$- and 3$\gamma$-decays, 
in which their ratio is taken as a free parameter.
This ratio is nothing but the pick-off annihilation probability of beam-OFF events, given by fitting the time spectra.
The lifetime of {\it o-}Ps decreases from 142~ns to approximately 131~ns due to this effect (the pick-off annihilation probability is about 8\%).
Relative differences of the two pick-off annihilation probabilities determined using these different methods are between $1$\% and $17$\% at the different frequencies, 
and are assigned as a systematic uncertainty to $S/N$.
These errors are propagated to obtained cross-sections and then to the three fitting parameters.

\begin{table}
\begin{center}
\caption{Summary of the systematic errors.\label{tab:sys}}
\begin{tabular}{lccc} \hline
\multicolumn{1}{c}{Source} & $\Delta^{\rm Ps}_{\rm HFS}$   & $\Gamma_{\rm p\text{--}Ps}$ & $A$ \\ \hline
Power estimation           & $430$~ppm   & $10.0$~\%           & $7.2$~\%   \\
Stark effect               & $460$~ppm   & $-$                & $-$        \\
Monte Carlo simulation     & $280$~ppm   & $5.5$~\%           & $3.0$~\%    \\ \hline
Total                      & $540$~ppm   & $11.4$~\%            & $7.8$~\% \\ \hline
\end{tabular}
\end{center}
\end{table}

\begin{table}[h]
\begin{center}
\caption{Summary of results. The first error is statistical and the second is systematic.\label{tab:result}}
\begin{tabular}{lcc}
\hline
\multicolumn{1}{c}{Parameter}         & This experiment  & Theory  \\ \hline
 $\Delta^{\rm Ps}_{\rm HFS}$~[GHz]    & $203.39 ^{+0.15}_{-0.14}\pm 0.11$  & $203.391\ 91(22)$ \\
 $\Gamma_{\rm p\text{--}Ps}$~[ns$^{-1}$]   & $11.2 ^{+1.9}_{-2.3}\pm1.3$   & $7.9894\ 76(13)$     \\
 $A$~[$\times10^{-8}\, {\rm s}^{-1}$] & $3.69 \pm 0.48\pm 0.29$ & $3.37$  \\ \hline 
\end{tabular}
\end{center}
\end{table}

The systematic errors discussed above are independent, and are therefore summed quadratically to calculate the total systematic error.
The obtained fitting parameters are listed in Table~\ref{tab:result}.
This is the first direct measurement of both $\Delta^{\rm Ps}_{\rm HFS}$ and $\Gamma_{\rm p\text{--}Ps}$. 
These all are consistent with the theoretical predictions~\cite{HFS-THEORY}\cite{PARA-THEORY}\cite{A-THEORY}.

\section{Discussion}
In this paper, we firstly demonstrate that $\Delta^{\rm Ps}_{\rm HFS}$ can be directly determined with the millimeter-wave spectroscopy.
A conventional method is using the Zeeman shifted levels caused by a static magnetic field.
In a static magnetic field ($\sim$1~T), one of the {\it o-}Ps states are mixed with {\it p-}Ps 
and the energy level of the mixed {\it o-}Ps state rises by about 3~GHz compared to the original state.
This Zeeman splitting can be precisely measured by an RF, being scanned by strength of the magnetic field.
The value of $\Delta^{\rm Ps}_{\rm HFS}$ is calculated via the Breit-Rabi formula~\cite{Breit-Rabi}.

In the 1970s and 1980s, 
the measurements with the Zeeman effect reached accuracies of ppm level~\cite{HFS-EXP}.
It should be noted that the obtained $\Delta^{\rm Ps}_{\rm HFS}$ significantly differs by 13~ppm from theoretical predictions calculated in the 2000s~\cite{HFS-THEORY}.
It may be due to underestimated systematic errors in the previous measurement.
For example, non-uniformity of the static magnetic field is a candidate of the systematic uncertainty.
Some independent experiments 
(using quantum interference~\cite{HFS-OSCI}, optical lasers~\cite{HFS-LASER}, and a new method using a precise magnetic field~\cite{HFS-PREC}) 
have been performed. 
All of them are measurements using the Zeeman intervals.
It is of great importance to re-measure $\Delta^{\rm Ps}_{\rm HFS}$ using a method totally different from the previous experiments.
Determination of $\Delta^{\rm Ps}_{\rm HFS}$ by directly measuring the transition from free {\it o-}Ps to {\it p-}Ps is a complementary approach to the measurements using the Zeeman effect.

We now discuss three improvements to achieve accuracy of 10~ppm level for $\Delta^{\rm Ps}_{\rm HFS}$:
\begin{enumerate}

\item Using a high-intensity positron beam (intensity of 7$\times10^7$e$^{+}$/s is available in KEK~\cite{WADA}) would increase the statistics by four orders of magnitude
because only a few~kHz Ps is formed inside the Fabry-P\'erot cavity using the $^{22}$Na source.
The statistical error becomes smaller than 10~ppm.

\item Positronium will be formed in vacuum using an efficient Ps converter (conversion efficiency is around 20--50\%~\cite{PSCONV}).
The Stark effect ($460$~ppm at 1~atm) and non-thermalization effect of Ps (about $10$--$20$~ppm) can be eliminated.
Since there is no pick-off annihilation in vacuum, $S/N$ will be also improved significantly by a factor of two.

\item Using a megawatt (MW) class gyrotron~\cite{MW}\cite{SAKAMOTO} would enable us to precisely (better than $0.3$~\%) monitor the real power with a calorimeter.
The present accuracy ($20$~\%) of the power estimation is mainly limited by uncertainty of the effective power in the Fabry-P\'erot cavity.
The systematic error due to the power can be better than 10~ppm.

\end{enumerate}

All these improvements have been technically achieved in the region of positron science and millimeter-wave technology.
Therefore, we can further investigate the disagreement of 4.0 standard deviations between measured $\Delta^{\rm Ps}_{\rm HFS}$ and QED theory 
with the direct measurement firstly reported in this paper.

\section{Conclusion}
We firstly measured the Breit-Wigner resonance of the transition from {\it o-}Ps to {\it p-}Ps with the frequency-tunable millimeter-wave system.
Both $\Delta^{\rm Ps}_{\rm HFS}$ and $\Gamma_{\rm p\text{--}Ps}$ of free Ps were directly and firstly determined through this resonance.
We pointed out the displacement of $\Delta^{\rm Ps}_{\rm HFS}$ between the previous experiments using the Zeeman effect and the theoretical calculations
can be tested by improving accuracy of this direct experiment.
Both direct and indirect measurements would be required to conclusively solve the long-standing problem on the ground-state hyperfine structure of Ps.

\section*{Acknowledgements}
We would like to express our sincere gratitude to Dr. Daniel Jeans for useful discussions.
This experiment is a joint research between Research Center for Development of Far-Infrared Region in University of Fukui and the University of Tokyo.
This research is supported by JSPS KAKENHI Grant Number 20340049, 22340051, 20840010, 21360167, 23740173, 24840011, 25800129, and 11J07131.


\vskip3pc

\begin{thebibliography}{100}

\bibitem{Ps-REV}
A. Rich, Rev. Mod. Phys. {\bf 53}, 127 (1981);
S. G. Karshenboim, Phys. Rep. {\bf 422}, 1 (2005).
T. Namba, Prog. Theor. Exp. Phys. 04D003 (2012).

\bibitem{LIFE-ORTH}
S. Asai, S. Orito, and N. Shinohara, Phys. Lett. B {\bf 357}, 475 (1995); O. Jinnouchi, S. Asai, and T. Kobayashi, {\it ibid.}~{\bf 572}, 117 (2003); Y. Kataoka, S. Asai, and T. Kobayashi, {\it ibid.}~{\bf 671}, 219 (2009);
R. S. Vallery, P. W. Zitzewitz, and D. W. Gidley, Phys. Rev. Lett. {\bf 90}, 203402 (2003). 

\bibitem{LIFE-PARA}
A. H. Al-Ramadhan and D. W. Gidley, Phys. Rev. Lett. {\bf 72}, 1632 (1994). 

\bibitem{DNP-NMR}
R. Ikeda, \etal, Plasma and Fusion Research \textbf{9}, 1206058 (2014);
M. Toda, \etal, J. Magn. Reson. \textbf{225}, 1 (2012);
K. J. Pike, \etal, J. Magn. Reson. \textbf{215}, 1 (2012).

\bibitem{HFS-DIRECT}
T. Yamazaki, \etal, Phys. Rev. Lett. \textbf{108}, 253401 (2012).

\bibitem{FU-CW-G1}
Y. Tatematsu, \etal, J. Infrared Milli. Terahz Waves {\bf 33}, 292 (2012).

\bibitem{CST}
CST Microwave Studio 2011, CST Computer Simulation Technology AG, http://www.cst.com

\bibitem{OPTICS}
A. Miyazaki, \etal, J. Infrared Milli. Terahz Waves {\bf 35}, 1, 91 (2014).

\bibitem{HFS-FIRST}
M. Deutsch and S. C. Brown, Phys. Rev. {\bf 85}, 1047 (1952). 

\bibitem{PS-FORM}
S. Marder, \etal, Phys. Rev. {\bf 103}, 1258 (1956);
W.B. Teutch and V.W. Hughes, Phys. Rev. {\bf 103}, 1266 (1956).

\bibitem{SLOW-POS}
K. Iwata, R.G. Greaves, T.J. Murphy, M.D. Tinkle, and C.M. Surko, Phys. Rev. A {\textbf 51}, 473 (1995);
K. Iwata, R.G. Greaves, and C.M. Surko, Phys. Rev. A {\textbf 55}, 3586 (1997).

\bibitem{PHD-MIYAZAKI}
A. Miyazaki, Ph.D. thesis, the University of Tokyo, 2014, p.~51.

\bibitem{ROTATION}
D.R. Lide, Jr. and D.E. Mann, J. Chem. Phys. {\textbf 29}, 4, 914 (1958);
D.R. Lide, J. Chem. Phys. {\textbf 33}, 5, 914 (1960).

\bibitem{PICK-OFF}
K. Wada, \etal, Eur. Phys. J. D {\bf 66}, 108 (2012).

\bibitem{GEANT4}
A. Agostinelli \etal, Nucl. Instrum. Method Phys. Res., Sect. A {\bf 506}, 250 (2003).

\bibitem{HFS-EXP}
A. P. Mills, Phys. Rev. A {\bf 27}, 262 (1983);
M. W. Ritter, \etal, Phys. Rev. A {\bf 30}, 1331 (1984). 

\bibitem{DOPPLER}
M. Skalsey, \etal, Phys. Rev. A, {\bf 67}, 022504 (2003).

\bibitem{THERM-OKINAWA}
S. Asai, \etal, AIP Conf. Proc. {\bf 1037}, 43 (2008).

\bibitem{HFS-PREC}
A. Ishida, \etal, Phys. Lett. B, {\bf 734}, 338 (2014).


\bibitem{HFS-THEORY}
B. A. Kniehl, and A. A. Penin, Phys. Rev. Lett. {\bf 85}, 5094 (2000); 
K. Melnikov and A. Yelkhovsky, Phys. Rev. Lett. {\bf 86}, 1498 (2001); 
R. J. Hill, Phys. Rev. Lett. {\bf 86}, 3280 (2001);
M. Baker, \etal, Phys. Rev. Lett. {\bf 112}, 120407 (2014);
G. S. Adkins and R. N. Fell, Phys. Rev. A {\bf 89}, 052518 (2014).

\bibitem{PARA-THEORY}
B. A. Kniehl, and A. A. Penin, Phys. Rev. Lett. {\bf 85}, 1210 (2000); 
K. Melnikov and A. Yelkhovsky, Phys. Rev. D. {\bf 62}, 116003 (2000).

\bibitem{A-THEORY}
P. Wallyn, \etal, Astrophys. J. {\bf 465}, 473 (1996).

\bibitem{Breit-Rabi}
M. L. Lewis and V. W. Hughes, Phys. Rev. A {\bf 8}, 625 (1973);
J. M. Anthony and K. J. Sebastian, Phys. Rev. A {\bf 49}, 1 (1994).

\bibitem{HFS-OSCI} 
V. G. Baryshevsky, O. N. Metelitsa, and V. V. Tikhomirov, J. Phys. B: At. Mol. Opt. Phys. \textbf{22}, 2835 (1989);
V. G. Baryshevsky, \etal, Phys. Lett. A \textbf{136}, 428 (1989);
S. Fan, C. D. Beling, and S. Fung, Phys. Lett. A \textbf{216}, 129 (1996);
Y. Sasaki, \etal, Phys. Lett. B \textbf{697}, 121 (2011).

\bibitem{HFS-LASER}
D. B. Cassidy, \etal, Phys. Rev. Lett. \textbf{109}, 073401 (2012).

\bibitem{WADA}
K. Wada, \etal, Eur. Phys. J B {\bf 66}, 37 (2012);

\bibitem{PSCONV}
P. J. Schultz and K. G. Lynn, Rev. Mod. Phys. {\bf 60}, 701, (1988);
P. Sferlazzo, \etal, Rev. Phys. Lett. {\bf 60}, 6, (1988).
A. P. Mills Jr., (1981). Experimentation with Low-Energy Positron Beams, 
Positron Solid State Physics: Proceedings of the International School of Physics "Enrico Fermi", Course LXXXIII, Varenna, 14-24 July (pp. 432-509), Elsevier Sci. Ltd.

\bibitem{MW}
G. Dammertz, \etal, IEEE Trans. Plas. Sci. {\bf 27}, 2 (1999).

\bibitem{SAKAMOTO}
K. Sakamoto, \etal, Nucl. Fusion {\bf 49}, 095019 (2009).







\end{thebibliography}
\end{document}